\documentclass[prl,twocolumn,showpacs,superscriptaddress,preprintnumbers,amsmath,amssymb]{revtex4}

\usepackage{graphicx, color}
\usepackage{dcolumn}
\usepackage{bm}

\newcommand\diag{\operatorname{diag}}

\newcommand\p{\partial}

\newcommand\T{{\rm T}}

\newcommand\lm{{\lambda} }

\usepackage{amsmath}	
\begin{document}

\title{The chiral magnetic effect in heavy-ion collisions\\ 
from event-by-event anomalous hydrodynamics}
\author{Yuji~Hirono}
\email{yuji.hirono@stonybrook.edu}
\affiliation{
Department of Physics and Astronomy, Stony Brook University,
Stony Brook, New York 11794-3800, USA
}

\author{Tetsufumi~Hirano}
\affiliation{
Department of Physics, Sophia University, Tokyo 102-8554, Japan
}

\author{Dmitri~E.~Kharzeev}
\affiliation{
Department of Physics and Astronomy, Stony Brook University,
Stony Brook, New York 11794-3800, USA
}
\affiliation{
Department of Physics, Brookhaven National Laboratory, Upton, New York 11973-5000, USA
}
\date{\today}

\begin{abstract}
The (3+1)D relativistic hydrodynamics with chiral anomaly is used to obtain
a quantitative description of the chiral magnetic effect (CME) in
 heavy-ion collisions. We find that the charge-dependent hadron
 azimuthal correlations are sensitive to the CME, and that the experimental
observations are consistent with the presence of the effect. 
\end{abstract}

\maketitle

The experimental study of charge-dependent hadron azimuthal correlations
in heavy-ion collisions at RHIC \cite{Abelev:2009ac,Abelev:2009ad} and
LHC \cite{Abelev:2012pa} revealed a signal qualitatively consistent with
the separation of electric charge predicted \cite{Kharzeev:2004ey} as a
signature of local ${\cal P}$- and ${\cal CP}$-odd fluctuations in QCD
matter. The subsequent studies
\cite{Kharzeev:2007tn,Kharzeev:2007jp,Fukushima:2008xe} improved the 
 theoretical understanding of the underlying phenomenon -- the
separation of electric charge in the quark-gluon plasma induced by the
chirality imbalance in the presence of background magnetic field, or the
``chiral magnetic effect'' (CME), see Ref.~\cite{Kharzeev:2013ffa} for a
review and additional references. The existence of CME has been
confirmed in first-principle lattice QCD$\times$QED simulations
\cite{Buividovich:2009wi,Abramczyk:2009gb,Yamamoto:2011gk,Bali:2014vja}. By
holographic methods, the CME has also been found to persist at strong
coupling \cite{Yee:2009vw}, in accord with its non-dissipative,
topologically protected nature.

Because the macroscopic behavior of matter at strong coupling is
described by hydrodynamics, it is natural to address the question of the
existence of CME within the framework of fluid dynamics. Son and Surowka
\cite{Son:2009tf} showed that the CME indeed is an integral part of
relativistic hydrodynamics, and moreover its strength as fixed by the
second law of thermodynamics is consistent with the field-theoretical
prediction. The CME current in the hydrodynamic regime is carried by a
novel collective gapless excitation, the chiral magnetic wave
\cite{Kharzeev:2010gd,Burnier:2011bf}. Conformal anomalous hydrodynamics
at second order in the derivative expansion has been formulated in \cite{Kharzeev:2011ds}. 

\begin{figure}[b]
\begin{center}
\includegraphics[width=80mm]{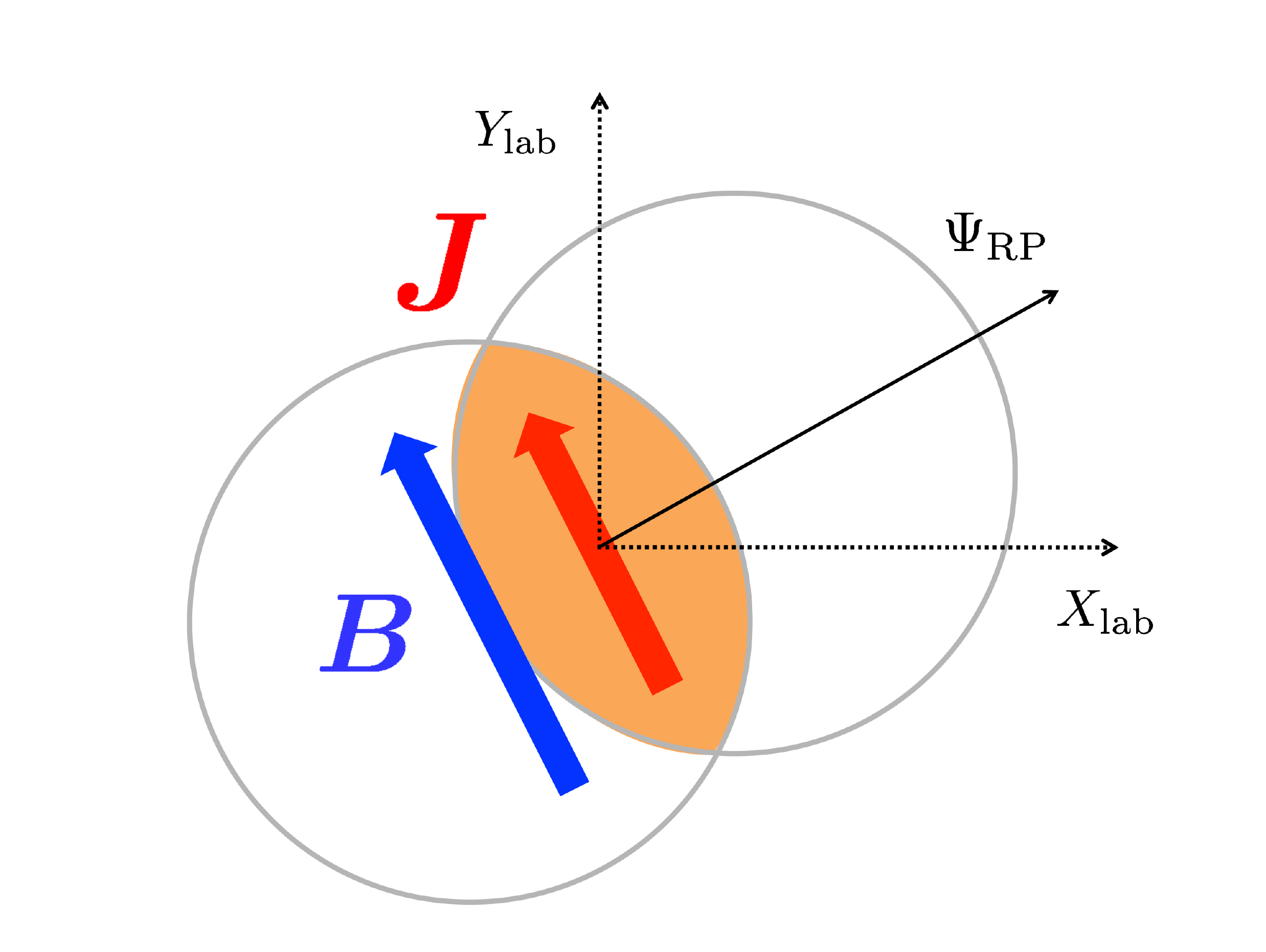}
\end{center}
\caption{
Schematic picture of a heavy-ion collision event in the transverse plane. 
$\bm B$ indicates the magnetic field, and $\bm J$ is the 
electric current induced by the chiral magnetic effect in the case of a
positive initial axial charge density. 
The direction of $\bm J$ is flipped if the initial axial charge is
 negative. 
}
\label{fig:coord}
\end{figure}

Relativistic hydrodynamics with fluctuating initial conditions proved
very successful in explaining the bulk of RHIC and LHC data (see
Refs.~\cite{Hirano:2012kj,Heinz:2013th,Petersen:2014yqa} for recent
reviews). It is thus
appropriate to rely on this approach also for describing the CME in
heavy-ion collisions, by using hydrodynamics with the terms induced by
chiral anomaly and magnetic fields -- so-called anomalous hydrodynamics,
or chiral magnetohydrodynamics (CMHD). The first pilot numerical study
of anomalous hydrodynamics was performed in Ref. \cite{Hongo:2013cqa},
where the effects of anomaly on the charge-dependent elliptic flow
were investigated. Nevertheless, a fully quantitative description of the
experimental data on charge-dependent azimuthal correlations has not
been performed until now. In this Letter, we perform such a study basing
on three-dimensional ideal CMHD with event-by-event fluctuations in the
initial conditions. 
 
Let us first describe the measured experimental observable
\cite{Voloshin:2004vk} sensitive to the CME that we aim to describe. The
azimuthal-angle distribution of observed particles reads
\begin{equation}
\begin{split}
\frac{d N^\alpha}{d \phi} &\propto 
1 + 2 v_1^\alpha \cos (\phi - \Psi_{\rm RP}) + 2 a_1^\alpha \sin (\phi - \Psi_{\rm
RP}) \\ 
 & \quad\quad + \sum_{n > 1} 2 v_n^\alpha \cos n (\phi - \Psi_{n} ), 
\end{split}
\label{eq:dn}
\end{equation}
where $\alpha \in \{+,-\}$. 
The Fourier coefficients $v_n^\pm$ of the azimuthal-angle distribution characterize the shape of the produced
matter in momentum space. 
The component with $n=1$ is called the directed flow. 
In Eq.~(\ref{eq:dn}), we decomposed the directed flow into two
directions, along ($v_1^\alpha$) and perpendicular to ($a_1^\alpha$) the
reaction plane angle $\Psi_{\rm RP}$ (see Fig.~\ref{fig:coord}).

The coefficients $a_1^\pm$ are expected to exhibit the presence of CME \cite{Kharzeev:2004ey,Voloshin:2004vk}. 
Indeed, suppose that a heavy-ion collision creates a localized lump of a positive axial charge. Because of the presence of the out-of-plane magnetic field \cite{Kharzeev:2007jp}, the CME current is generated in the direction of the magnetic field, 
leading to the formation of a charge dipole . The collective expansion of matter translates this dipole into the charge-dependent flow quantified by $a_1^\pm$.
For a positive initial axial charge, $a_1^+ > 0$ and $a_1^- < 0$, since positive particles tend
to move along the direction of magnetic field and negative particles
move in the opposite direction. 
In contrast, if the initial axial charge is negative, 
the result is $a_1^+ < 0$ and $a_1^- > 0$ since the direction of
the electric current is reversed. 
The coefficients $a_1^\pm$ can thus indeed be affected by the CME.

However, $a_1^\pm$ in a single event is very sensitive to statistical
fluctuations -- we are thus forced to study the quantities averaged over
many events. Since there is no preferred sign of the initial axial charge, 
the event averages of $a_1^\pm$ vanish, 
$\langle a_1^+ \rangle = \langle a_1^- \rangle = 0$. 
We thus have to look at the fluctuations of $a_1^\pm$. 
The CME-based expectations for these quantities are the following: 
\begin{enumerate}
 \item The fluctuation $\langle \left( a_1^\pm \right)^2 \rangle$ would
       become larger in the presence of anomalous transport effects. 
 \item There would be an anti-correlation between $a_1^+$ and $a_1^-$, 
       so that $\langle a_1^+ a_1^- \rangle < 0$. 
\end{enumerate}

To quantify the strength of ${\cal P}$-odd fluctuations, the following observable has been proposed  \cite{Voloshin:2004vk} and measured by the STAR Collaboration \cite{Abelev:2009ad}: 
\begin{equation}
 \gamma_{\alpha\beta} \equiv 
\langle
\cos (
\phi^\alpha_1 + \phi^\beta_2 - 2 \Psi_{\rm RP} 
)
\rangle, 
\label{eq:gamma}
\end{equation}
where $\phi^\alpha_1$ and $\phi^\beta_2$ are the azimuthal angles of the
first and second particle, and $\alpha, \beta \in \{+,-\}$ indicate their 
charges. The physical meaning of the observable (\ref{eq:gamma}) becomes evident
if we re-write ($\Delta \phi^\alpha \equiv \phi^\alpha - \Psi_{\rm RP}$)
\begin{equation}
\begin{split}
 \langle \cos (\phi_1^\alpha + \phi_2^\beta - 2 \Psi_{\rm RP}) \rangle
= 
 \langle \cos \Delta \phi^\alpha_1 \cos \Delta \phi_2^\beta
\rangle  \\
- 
 \langle \sin \Delta \phi^\alpha_1 \sin \Delta \phi_2^\beta
\rangle \equiv 
\langle
v_1^\alpha v_1^\beta
\rangle
- 
\langle
a_1^\alpha a_1^\beta
\rangle. 
\end{split}
\label{eq:decomp}
\end{equation}
One can see that $\gamma_{++}$ and $\gamma_{--}$ are sensitive to 
$\langle (a^\pm_1)^2 \rangle$, and 
$\gamma_{+-}$ to  $\langle a_1^+ a_1^- \rangle$. Another merit of using (\ref{eq:decomp}) is that the reaction-plane-independent background effects 
are eliminated by the subtraction between 
$\langle (v_1)^2\rangle$ and $\langle (a_1)^2 \rangle$ 
\cite{Abelev:2009ad}. 

The equations of motion of anomalous hydrodynamics are given by
\begin{align}
 \p_\mu T^{\mu\nu} &= e F^{\nu \lm} j_{\lm}, \\
\p_\mu j^\mu &= 0, \\
\p_\mu j_5^\mu &= -C E_\mu B^\mu, 
\end{align}
where 
$
C \equiv \frac{N_c} {2 \pi^2} \sum_{f} q_f^2
$ is the anomaly constant, 
$E^\mu \equiv F^{\mu\nu} u_{\nu}$, $B^\mu \equiv \tilde F^{\mu\nu} u_{\nu}$
with $ \tilde F^{\mu\nu} = \frac{1}{2} \epsilon^{\mu\nu\alpha\beta}
F_{\alpha\beta}$. 
In this study, the electromagnetic fields are treated as background
fields. We consider three light flavors of quarks and for simplicity put
their electric charges $q_f = 1$; introducing real charges of quarks
would amount to a simple rescaling of charge densities in the equations
below. 
The fluid is assumed to be dissipationless (note that the CME current is non-dissipative),  
so the energy-momentum tensor, the electric and axial currents are
written as 
\begin{align}
 T^{\mu\nu} &= (\varepsilon +p)u^\mu u^\nu - p \eta^{\mu\nu}, \\ 
 j^\mu &= n u^\mu + \kappa_B B^\mu, \label{eq:cons-j}\\
 j_5^\mu &= n_5 u^\mu + \xi_B B^\mu,\label{eq:cons-j5}
\end{align}
where $\varepsilon$ is the energy density, $p$ is the hydrodynamic
pressure, $n$ and $n_5$ are electric and axial charge densities, 
$e \kappa_B \equiv  C \mu_5 ( 1 - \mu_5 n_5 / (\varepsilon+p))$ and 
$e \xi_B \equiv  C \mu ( 1 - \mu n / (\varepsilon +p))$ are transport
coefficients, 
and $\eta^{\mu\nu} \equiv \diag \{1,-1,-1,-1\}$ is the Minkowski metric. 
The second terms in Eq.~(\ref{eq:cons-j}) and Eq.~(\ref{eq:cons-j5})
correspond to the chiral magnetic effect and the chiral separation
effect, respectively. The values of $\kappa_B$ and $\xi_B$ are determined by requiring
that the entropy does not decrease \cite{Son:2009tf}, or from the condition of zero entropy production from anomalous currents \cite{Kharzeev:2011ds}, with an equivalent result.

Hydrodynamic equations should be augmented with an equation of state
(EOS). 
Here we use EOS of a massless ideal quark-gluon gas \footnote{In hydrodynamic models, it is now standard to
use an EOS calculated from lattice QCD. In the present calculations, we
need $\mu$ and $\mu_5$ dependence, which is not yet available from the lattice -- so here we use the EOS of an ideal  gas. 
},
$
\varepsilon = 3p
$, with 
\begin{equation}
\begin{split}
p\left(
T,\mu,\mu_5
\right) &= 
\frac{g_{\rm QGP} \pi^2}{90 }T^4 + \frac{N_c N_f}{6}\left( \mu^2
						     +\mu_5^2 \right)
T^2 \\
&\quad + \frac{N_c N_f}{12 \pi^2} 
\left(
 \mu^4 + 6 \mu^2 \mu_5^2 + \mu_5^4
\right), 
\label{eq:eosp}
\end{split}
\end{equation}
where $g_{\rm QGP} = g_{g} + \frac{7}{8}g_{q}$ is the number of
degrees of freedom with $g_{g} = (N_c^2 - 1) N_s$ and $g_{q} = 2 N_c N_s
N_f$; $N_c = 3$ and $N_f = 3$ are the numbers of colors and flavors and $N_s =2$ is the number of spin states for quarks and (transverse) gluons. 
The entropy, electric, and axial charge densities are obtained 
from Eq.~(\ref{eq:eosp}) as 
$s = \frac{\p p}{\p T}$, 
$n = \frac{\p p}{\p \mu}$, and 
$n_5 = \frac{\p p}{\p \mu_5}$. 

\begin{figure*}[htbp]
\begin{center}
\includegraphics[width=185mm]{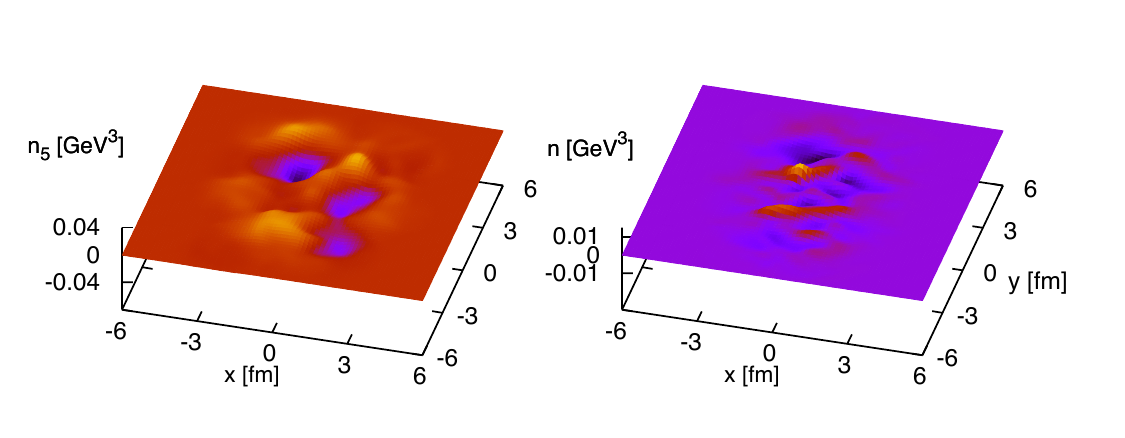}
\caption{
Axial charge (left) and electric charge (right) densities 
at $\tau = 1.5 \ {\rm fm}$ in the transverse plane ($\eta_s = 0$). 
}
\label{fig:n-n5}
\end{center}
\end{figure*}

In order to switch from hydrodynamic degrees of freedom to individual particles, 
we use the standard Cooper-Frye approach \cite{Cooper:1974mv}, 
on an isothermal freezeout hypersurface with $T_{\rm fo} = 160 \ {\rm MeV}$. 
We perform Monte-Carlo samplings of hadrons on the freezeout surface 
following the procedure described in detail in Ref.~\cite{Hirano:2012kj}. 

As a background electromagnetic field, we take $B_y$ to be 
($x$-axis is taken to be $\Psi_{\rm RP}$)
\begin{equation}
 e B_y( \tau, \eta_s ,  \bm x_\perp) = e B_0 \frac{b}{2R} \exp \left[
- \frac{x^2}{\sigma_x^2}
- \frac{y^2}{\sigma_y^2}
- \frac{\eta_s^2}{\sigma_{\eta_s}^2}
-\frac{\tau}{\tau_{\rm B}}
\right] , 
\label{eq:magnetic}
\end{equation}
where $\sigma_x$, $\sigma_y$, and $\sigma_{\eta_s}$ are the widths of
the field in $x$, $y$, and $\eta_s$ (space-time rapidity) directions, $\tau_B$ is the
decay time of the magnetic field, $R = 6.38 \ {\rm fm}$ is the radius
of a gold nucleus, and $b$ is the impact parameter. 
Other components of $\bm B$ and $\bm E$ were set to zero. 
We assume the strength of the magnetic field (\ref{eq:magnetic}) to be 
proportional to the impact parameter \cite{Kharzeev:2007jp}. 
The widths are chosen as  
$\sigma_{x} = 0.8 \left(R - \frac{b}{2}\right)$, 
$\sigma_{y} = 0.8 \sqrt{R^2 - (b/2)^2 }$,
and $\sigma_\eta = \sqrt{2}$, 
so that the fields are applied only in the region where matter exists.
We set $\tau_B = 3 \ {\rm fm}$ and $eB_0 = 0.5 {\rm GeV}^2$ for the
following calculations, which corresponds to $ e B_y(\tau_{\rm in}, 0, \bm
0) \sim (3 m_\pi)^2$. 

For the modeling of the initial axial charge density, we make an
extension to the Monte-Carlo (MC) Glauber model \cite{Alver:2008aq}. 
We assume that the initial axial charge density is generated by the
color flux tubes during the early moments of the heavy-ion collision
\cite{Kharzeev:2001ev,Lappi:2006fp}. In this picture, the parallel
chromo-electric $\bm E^a$ and chromo-magnetic $\bm B^a$ fields (with
strength $\sim Q_s^2/g$ determined by the saturation momentum $Q_s$)
create the axial charge through the chiral anomaly equation $\p_\mu
j^\mu_5 = \frac{g^2}{16 \pi^2} \bm E^a \cdot \bm B^a$.  
First, we assign $\pm 1$ for each binary collision randomly -- 
each sign corresponds to the sign of $\bm E^a \cdot \bm B^a$ for a color
flux tube where the color orientations of $\bm E^a$ and $\bm B^a$ change after each collision due to the color exchanges.   
Then, when we calculate the axial chemical potential for a particular
point in the transverse plane, we sum the signs over the collisions
occurring at that position and multiply it by a dimensionful constant $C_{\mu_5}$.

We estimate the parameter $C_{\mu_5}$ as follows. 
Integrating the chiral anomaly equation, we find the axial charge density at the time $\tau_{\rm in}$ at which the hydrodynamic evolution is initialized: 
\begin{equation}
n_5(\tau_{\rm in})
\sim 
\tau_{\rm in} \times 
\frac{g^2}{16 \pi^2}  
\bm E^a \cdot \bm B^a. 
\end{equation}
We take $
\tau_{\rm in} \sim 0.6 \ {\rm fm}  \simeq 3 \ {\rm GeV^{-1}}, 
$ 
and estimate the strength of the gluon fields from the saturation scale at RHIC as 
$ g | \bm E^a| \sim g | \bm B^a | \sim 1  \ {\rm GeV}^2$ \cite{Kharzeev:2000ph,Kharzeev:2001gp} 
$
n_5(\tau_{\rm in}) \sim \left( 0.4 \ {\rm GeV} \right)^3
$,
translating into $\mu_5$ is of the order of $0.1 \ {\rm GeV}$; we thus set 
 the coefficient $C_{\rm \mu_5} = 0.1 \ {\rm GeV}$.  From the values of $\mu_5 (\bm x_\T, \eta_{\rm s})$ and the entropy $s(\bm x_{\rm T},
\eta_{\rm s})$, we determine the initial $T$. 
The initial flow velocity and electric charge density are taken to be zero \footnote{
Charge separation can also occur before the onset of hydrodynamic
evolution. 
Here, by taking $n(\tau_{\rm in}, \eta_s, \bm x_{\rm T})=0$, we are
focusing on the charge separation in the hydro phase. 
The charge separation in the non-equilibrium gluonic matter is discussed
in \cite{Fukushima:2010vw}.
}.


We first generate an initial condition by the MC-Glauber model 
including the effect of fluctuating axial charge densities, at a fixed
impact parameter. 
Then we let the matter evolve according  to anomalous hydrodynamic
equations, getting a freezeout hypersurface. 
Sampling particles from the surface, we get the data for an event. 
We repeat this process to accumulate the data for many events. 
We performed the simulations for three impact parameters, $b=7.2, 9.0,
11.2 \ {\rm fm}$ 
(20-30\%, 30-40\%, 50-60\% in centrality). 
For each impact parameter, we calculated 100k, 150k, and 250k events for
both of anomalous and non-anomalous cases. 
In the ``non-anomalous'' case, the anomaly constant is set to zero and
there is no CME. 

Let us first look at how the charge profile is affected by the anomalous
transport. 
Figure \ref{fig:n-n5}  shows the axial charge density and electric charge
densities in the transverse plane ($\eta_s=0$) for an event. 
As is evident from the right figure, the CME leads to a non-trivial 
electric charge profile. 
(If we switch off the anomalous transport
effect, the electric charge density is zero throughout the evolution.)
One can see a number of charge dipoles formed in areas where there was a non-zero axial
charge. 
The direction of a dipole is determined by the sign of the axial charge
of density.

\begin{figure}[tbp]
\begin{center}
\includegraphics[width=86mm]{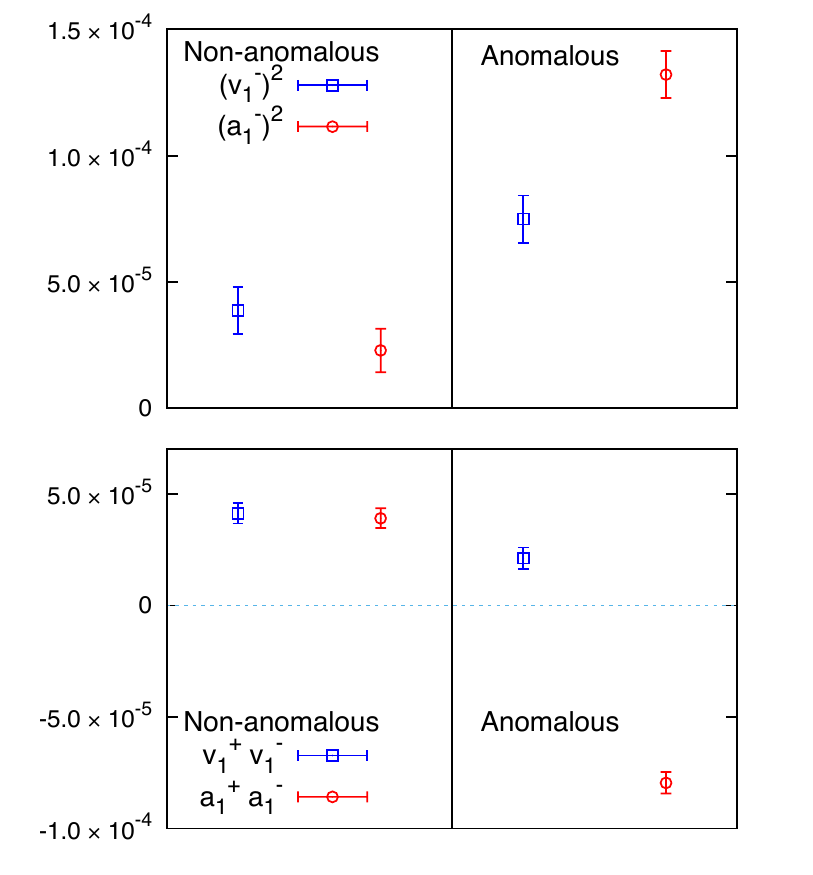}
\end{center}
\caption{
$\langle \left(v^{-}_1\right)^2 \rangle$, $\langle
 \left(a^-_1\right)^2 \rangle$ (upper figure), 
$\langle v_1^+ v_1^- \rangle$, and $\langle a_1^+ a_1^- \rangle$ (lower
 figure) 
for anomalous and non-anomalous cases 
at $b=7.2 \ {\rm fm}$ (20-30\% in centrality).
}
\label{fig:vapm2}
\end{figure}

Next, let us turn to the charge-dependent correlation functions. 
The correlations $\gamma_{++}$($\gamma_{--}$) and $\gamma_{+-}$ are
calculated via 
\begin{align}
 \gamma_{\alpha \alpha} &\equiv 
\left<
\frac{1}{_{M_\alpha} {\rm P}_{2} } \sum_{<i,j> \in S_\alpha } 
\cos (\phi_i + \phi_j
- 2 \Psi_{\rm RP})
\right>, \\
 \gamma_{\alpha\beta} &\equiv 
\left<
\frac{1}{M_\alpha M_\beta} \sum_{i \in S_\alpha }\sum_{j \in S_\beta}
\cos (\phi_i + \phi_j
- 2 \Psi_{\rm RP})
\right>, 
\end{align}
where $\alpha \neq \beta$, $S_\alpha$ is the set of particles with
charge $\alpha$, $<i,j> \in S$ in the summation symbol indicate the pair
of particles in a particle set $S$, and the overall $\langle \cdots \rangle$
means the event average. 
Similarly, we separately evaluate $\left(v_1^\alpha\right)^2$ and
$\left(a_1^\alpha\right)^2$ by
\begin{align}
\langle 
\left(v^\alpha_1\right)^2
\rangle
&\equiv
\left<
\frac{1}{_{M_\alpha} {\rm P}_2} 
\sum_{<i,j> \in S_\alpha}
\cos \Delta \phi_i^\alpha 
\cos \Delta \phi_j^\alpha 
\right>, \\
\langle 
\left(a^\alpha_1\right)^2
\rangle
&\equiv
\left<
\frac{1}{_{M_\alpha} {\rm P}_2} 
\sum_{<i,j> \in S_\alpha}
\sin \Delta \phi_i^\alpha
\sin \Delta \phi_j^\alpha
\right>.
\end{align}

Upper figure of Fig.~\ref{fig:vapm2} shows $\langle \left(v_1^-\right)^2 \rangle$
and $\langle \left(a_1^-\right)^2 \rangle$ in anomalous and
non-anomalous cases at $b=7.2 \ {\rm fm}$ (20-30\% in centrality). 
As shown in the figure, the fluctuation of $v_1$ becomes large in the
presence of CME, and 
the fluctuation of $a_1$ becomes even larger. 
The observable $\gamma_{--}$ is the difference between $v_1$ and $a_1$
fluctuations, so it becomes negative, which is consistent with the
experiment. 
The magnitude of $\gamma_{--}$ is also comparable to the values measured
in STAR. 
The large value of $\langle a_1^2 \rangle$ is in line with the expectation
from the CME. 

In the lower figure of Fig.~\ref{fig:vapm2}, we show the values of $\langle v_1^+ v_1^- \rangle$
and $\langle a_1^+ a_1^- \rangle$. 
When there is no anomalous transport, the values for 
$\langle v_1^+ v_1^- \rangle$ and $\langle a_1^+ a_1^- \rangle$ are
positive and comparable. 
When anomalous transport effects are turned on, 
$\langle a_1^+ a_1^- \rangle$ takes a negative value. 
This means the anti-correlation between $a_1^+$ and
$a_1^-$, which is expected from the CME.

\begin{figure}[tbp]
\begin{center}
\includegraphics[width=86mm]{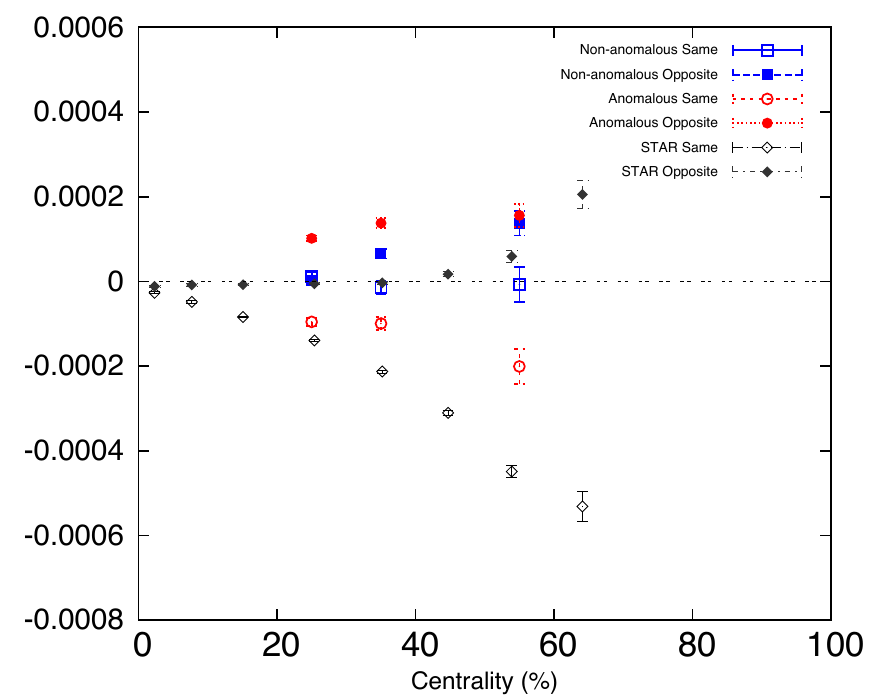}
\end{center}
\caption{
Centrality dependence of $\gamma_{++}$ and $\gamma_{+-}$ as well as the
 experimental data from STAR \cite{Abelev:2009ad}. 
}
\label{fig:cent}
\end{figure}

In Fig.~\ref{fig:cent}, the centrality dependence of the
observables $\gamma_{++}$ and $\gamma_{+-}$ as well as the experimental
values from STAR \cite{Adamczyk:2013hsi} are plotted. 
The values of the same-charge combination $\gamma_{++}$ 
are negative and the magnitude increases as a function of centrality
in the anomalous case, 
while the values are consistent with zero in the non-anomalous case. 
As for the opposite charge combination $\gamma_{+-}$, the values are
positive and are larger in peripheral collisions in the anomalous case. 
The positiveness comes from negative values of 
$\langle a_1^+ a_1^- \rangle$, which indicates the anti-correlation
between $a_1^+$ and $a_1^-$. 
At two peripheral centralities, $\gamma_{+-}$ is positive even for the
non-anomalous case.

Let us comment on the implications of this work to the possible  
background effects discussed in the literature. It has been pointed out
that  the observed values of $\gamma_{\alpha\beta}$ may be reproduced by
other effects unrelated to anomalous transport, including transverse
momentum conservation and flow \cite{Bzdak:2010fd,Pratt:2010gy}, charge
conservation and flow \cite{Schlichting:2010na}, 
or cluster particle correlations \cite{Wang:2009kd}. 
All of these background effects originate from multi-particle
correlations. 
Since neither local charge conservation nor
transverse momentum conservation are imposed in our current sampling
procedure, 
the computed correlators do not originate from these multi-particle
correlations - therefore, we expect the measured correlators to reflect the
presence of CME, even though of course we cannot exclude the possible
contributions from backgrounds.

We believe that event-by-event anomalous hydrodynamics is an appropriate tool for 
the quantitative test of the CME in heavy-ion collisions. 
However, it is still in the early stage and 
there is a lot of room for improvement in the current model. 
Possible future improvements include: {\it i}) event-by-event
fluctuations of magnetic fields \cite{Bzdak:2011yy, Bloczynski:2012en};
{\it ii}) refinement of the modeling of initial axial charge density,
possibly based on the numerical solution of Yang-Mills equations; {\it
iii}) dynamical electromagnetic fields; {\it i{\rm v}}) dissipative
effects, including shear and bulk viscosities, Ohmic conductivity, etc. 
In parallel with those improvements of the model, 
we plan to perform  a more detailed analysis of $\gamma_{\alpha\beta}$, including 
the dependence on transverse momentum, rapidity, collision energy and the nuclear species.

To summarize, we have performed a quantitative study of charge-dependent
azimuthal hadron correlations in heavy-ion collisions using anomalous
relativistic hydrodynamics. Our results indicate that these correlations
are sensitive to the chiral magnetic effect, and that the pattern and
the magnitude of measured correlations are in accord with anomalous
hydrodynamics -- even though the theoretical uncertainties are still
substantial.

\begin{acknowledgements}
The authors are grateful to M.~Hongo, K.~Murase, S.~Schlichting, 
Y.~Tachibana, and Y.~Yin for useful discussions. 
This work was supported in part by the U.S. Department of Energy under Contracts No.
DE-FG-88ER40388 and DE-AC02-98CH10886. 
Y.~H. is supported by JSPS Research Fellowships for Young Scientists. 
The work of T.~H. was supported by JSPS KAKENHI Grants No. 25400269.
\end{acknowledgements}

\bibliography{refs}

\begin{thebibliography}{36}
\expandafter\ifx\csname natexlab\endcsname\relax\def\natexlab#1{#1}\fi
\expandafter\ifx\csname bibnamefont\endcsname\relax
  \def\bibnamefont#1{#1}\fi
\expandafter\ifx\csname bibfnamefont\endcsname\relax
  \def\bibfnamefont#1{#1}\fi
\expandafter\ifx\csname citenamefont\endcsname\relax
  \def\citenamefont#1{#1}\fi
\expandafter\ifx\csname url\endcsname\relax
  \def\url#1{\texttt{#1}}\fi
\expandafter\ifx\csname urlprefix\endcsname\relax\def\urlprefix{URL }\fi
\providecommand{\bibinfo}[2]{#2}
\providecommand{\eprint}[2][]{\url{#2}}

\bibitem[{\citenamefont{Abelev et~al.}(2009)}]{Abelev:2009ac}
\bibinfo{author}{\bibfnamefont{B.}~\bibnamefont{Abelev}} \bibnamefont{et~al.}
  (\bibinfo{collaboration}{STAR Collaboration}),
  \bibinfo{journal}{Phys.Rev.Lett.} \textbf{\bibinfo{volume}{103}},
  \bibinfo{pages}{251601} (\bibinfo{year}{2009}), \eprint{0909.1739}.

\bibitem[{\citenamefont{Abelev et~al.}(2010)}]{Abelev:2009ad}
\bibinfo{author}{\bibfnamefont{B.}~\bibnamefont{Abelev}} \bibnamefont{et~al.}
  (\bibinfo{collaboration}{STAR Collaboration}), \bibinfo{journal}{Phys.Rev.}
  \textbf{\bibinfo{volume}{C81}}, \bibinfo{pages}{054908}
  (\bibinfo{year}{2010}), \eprint{0909.1717}.

\bibitem[{\citenamefont{Abelev et~al.}(2013)}]{Abelev:2012pa}
\bibinfo{author}{\bibfnamefont{B.}~\bibnamefont{Abelev}} \bibnamefont{et~al.}
  (\bibinfo{collaboration}{ALICE Collaboration}),
  \bibinfo{journal}{Phys.Rev.Lett.} \textbf{\bibinfo{volume}{110}},
  \bibinfo{pages}{012301} (\bibinfo{year}{2013}), \eprint{1207.0900}.

\bibitem[{\citenamefont{Kharzeev}(2006)}]{Kharzeev:2004ey}
\bibinfo{author}{\bibfnamefont{D.}~\bibnamefont{Kharzeev}},
  \bibinfo{journal}{Phys.Lett.} \textbf{\bibinfo{volume}{B633}},
  \bibinfo{pages}{260} (\bibinfo{year}{2006}), \eprint{hep-ph/0406125}.

\bibitem[{\citenamefont{Kharzeev and Zhitnitsky}(2007)}]{Kharzeev:2007tn}
\bibinfo{author}{\bibfnamefont{D.}~\bibnamefont{Kharzeev}} \bibnamefont{and}
  \bibinfo{author}{\bibfnamefont{A.}~\bibnamefont{Zhitnitsky}},
  \bibinfo{journal}{Nucl.Phys.} \textbf{\bibinfo{volume}{A797}},
  \bibinfo{pages}{67} (\bibinfo{year}{2007}), \eprint{0706.1026}.

\bibitem[{\citenamefont{Kharzeev et~al.}(2008)\citenamefont{Kharzeev, McLerran,
  and Warringa}}]{Kharzeev:2007jp}
\bibinfo{author}{\bibfnamefont{D.~E.} \bibnamefont{Kharzeev}},
  \bibinfo{author}{\bibfnamefont{L.~D.} \bibnamefont{McLerran}},
  \bibnamefont{and} \bibinfo{author}{\bibfnamefont{H.~J.}
  \bibnamefont{Warringa}}, \bibinfo{journal}{Nucl.Phys.}
  \textbf{\bibinfo{volume}{A803}}, \bibinfo{pages}{227} (\bibinfo{year}{2008}),
  \eprint{0711.0950}.

\bibitem[{\citenamefont{Fukushima et~al.}(2008)\citenamefont{Fukushima,
  Kharzeev, and Warringa}}]{Fukushima:2008xe}
\bibinfo{author}{\bibfnamefont{K.}~\bibnamefont{Fukushima}},
  \bibinfo{author}{\bibfnamefont{D.~E.} \bibnamefont{Kharzeev}},
  \bibnamefont{and} \bibinfo{author}{\bibfnamefont{H.~J.}
  \bibnamefont{Warringa}}, \bibinfo{journal}{Phys.Rev.}
  \textbf{\bibinfo{volume}{D78}}, \bibinfo{pages}{074033}
  (\bibinfo{year}{2008}), \eprint{0808.3382}.

\bibitem[{\citenamefont{Kharzeev}(2014)}]{Kharzeev:2013ffa}
\bibinfo{author}{\bibfnamefont{D.~E.} \bibnamefont{Kharzeev}},
  \bibinfo{journal}{Prog.Part.Nucl.Phys.} \textbf{\bibinfo{volume}{75}},
  \bibinfo{pages}{133} (\bibinfo{year}{2014}), \eprint{1312.3348}.

\bibitem[{\citenamefont{Buividovich et~al.}(2009)\citenamefont{Buividovich,
  Chernodub, Luschevskaya, and Polikarpov}}]{Buividovich:2009wi}
\bibinfo{author}{\bibfnamefont{P.}~\bibnamefont{Buividovich}},
  \bibinfo{author}{\bibfnamefont{M.}~\bibnamefont{Chernodub}},
  \bibinfo{author}{\bibfnamefont{E.}~\bibnamefont{Luschevskaya}},
  \bibnamefont{and}
  \bibinfo{author}{\bibfnamefont{M.}~\bibnamefont{Polikarpov}},
  \bibinfo{journal}{Phys.Rev.} \textbf{\bibinfo{volume}{D80}},
  \bibinfo{pages}{054503} (\bibinfo{year}{2009}), \eprint{0907.0494}.

\bibitem[{\citenamefont{Abramczyk et~al.}(2009)\citenamefont{Abramczyk, Blum,
  Petropoulos, and Zhou}}]{Abramczyk:2009gb}
\bibinfo{author}{\bibfnamefont{M.}~\bibnamefont{Abramczyk}},
  \bibinfo{author}{\bibfnamefont{T.}~\bibnamefont{Blum}},
  \bibinfo{author}{\bibfnamefont{G.}~\bibnamefont{Petropoulos}},
  \bibnamefont{and} \bibinfo{author}{\bibfnamefont{R.}~\bibnamefont{Zhou}},
  \bibinfo{journal}{PoS} \textbf{\bibinfo{volume}{LAT2009}},
  \bibinfo{pages}{181} (\bibinfo{year}{2009}), \eprint{0911.1348}.

\bibitem[{\citenamefont{Yamamoto}(2011)}]{Yamamoto:2011gk}
\bibinfo{author}{\bibfnamefont{A.}~\bibnamefont{Yamamoto}},
  \bibinfo{journal}{Phys.Rev.Lett.} \textbf{\bibinfo{volume}{107}},
  \bibinfo{pages}{031601} (\bibinfo{year}{2011}), \eprint{1105.0385}.

\bibitem[{\citenamefont{Bali et~al.}(2014)\citenamefont{Bali, Bruckmann,
  Endr{\"o}di, Fodor, Katz et~al.}}]{Bali:2014vja}
\bibinfo{author}{\bibfnamefont{G.}~\bibnamefont{Bali}},
  \bibinfo{author}{\bibfnamefont{F.}~\bibnamefont{Bruckmann}},
  \bibinfo{author}{\bibfnamefont{G.}~\bibnamefont{Endr{\"o}di}},
  \bibinfo{author}{\bibfnamefont{Z.}~\bibnamefont{Fodor}},
  \bibinfo{author}{\bibfnamefont{S.}~\bibnamefont{Katz}}, \bibnamefont{et~al.},
  \bibinfo{journal}{JHEP} \textbf{\bibinfo{volume}{1404}}, \bibinfo{pages}{129}
  (\bibinfo{year}{2014}), \eprint{1401.4141}.

\bibitem[{\citenamefont{Yee}(2009)}]{Yee:2009vw}
\bibinfo{author}{\bibfnamefont{H.-U.} \bibnamefont{Yee}},
  \bibinfo{journal}{JHEP} \textbf{\bibinfo{volume}{0911}}, \bibinfo{pages}{085}
  (\bibinfo{year}{2009}), \eprint{0908.4189}.

\bibitem[{\citenamefont{Son and Surowka}(2009)}]{Son:2009tf}
\bibinfo{author}{\bibfnamefont{D.~T.} \bibnamefont{Son}} \bibnamefont{and}
  \bibinfo{author}{\bibfnamefont{P.}~\bibnamefont{Surowka}},
  \bibinfo{journal}{Phys.Rev.Lett.} \textbf{\bibinfo{volume}{103}},
  \bibinfo{pages}{191601} (\bibinfo{year}{2009}), \eprint{0906.5044}.

\bibitem[{\citenamefont{Kharzeev and
  Yee}(2011{\natexlab{a}})}]{Kharzeev:2010gd}
\bibinfo{author}{\bibfnamefont{D.~E.} \bibnamefont{Kharzeev}} \bibnamefont{and}
  \bibinfo{author}{\bibfnamefont{H.-U.} \bibnamefont{Yee}},
  \bibinfo{journal}{Phys.Rev.} \textbf{\bibinfo{volume}{D83}},
  \bibinfo{pages}{085007} (\bibinfo{year}{2011}{\natexlab{a}}),
  \eprint{1012.6026}.

\bibitem[{\citenamefont{Burnier et~al.}(2011)\citenamefont{Burnier, Kharzeev,
  Liao, and Yee}}]{Burnier:2011bf}
\bibinfo{author}{\bibfnamefont{Y.}~\bibnamefont{Burnier}},
  \bibinfo{author}{\bibfnamefont{D.~E.} \bibnamefont{Kharzeev}},
  \bibinfo{author}{\bibfnamefont{J.}~\bibnamefont{Liao}}, \bibnamefont{and}
  \bibinfo{author}{\bibfnamefont{H.-U.} \bibnamefont{Yee}},
  \bibinfo{journal}{Phys.Rev.Lett.} \textbf{\bibinfo{volume}{107}},
  \bibinfo{pages}{052303} (\bibinfo{year}{2011}), \eprint{1103.1307}.

\bibitem[{\citenamefont{Kharzeev and
  Yee}(2011{\natexlab{b}})}]{Kharzeev:2011ds}
\bibinfo{author}{\bibfnamefont{D.~E.} \bibnamefont{Kharzeev}} \bibnamefont{and}
  \bibinfo{author}{\bibfnamefont{H.-U.} \bibnamefont{Yee}},
  \bibinfo{journal}{Phys.Rev.} \textbf{\bibinfo{volume}{D84}},
  \bibinfo{pages}{045025} (\bibinfo{year}{2011}{\natexlab{b}}),
  \eprint{1105.6360}.

\bibitem[{\citenamefont{Hirano et~al.}(2013)\citenamefont{Hirano, Huovinen,
  Murase, and Nara}}]{Hirano:2012kj}
\bibinfo{author}{\bibfnamefont{T.}~\bibnamefont{Hirano}},
  \bibinfo{author}{\bibfnamefont{P.}~\bibnamefont{Huovinen}},
  \bibinfo{author}{\bibfnamefont{K.}~\bibnamefont{Murase}}, \bibnamefont{and}
  \bibinfo{author}{\bibfnamefont{Y.}~\bibnamefont{Nara}},
  \bibinfo{journal}{Prog.Part.Nucl.Phys.} \textbf{\bibinfo{volume}{70}},
  \bibinfo{pages}{108} (\bibinfo{year}{2013}), \eprint{1204.5814}.

\bibitem[{\citenamefont{Heinz and Snellings}(2013)}]{Heinz:2013th}
\bibinfo{author}{\bibfnamefont{U.}~\bibnamefont{Heinz}} \bibnamefont{and}
  \bibinfo{author}{\bibfnamefont{R.}~\bibnamefont{Snellings}},
  \bibinfo{journal}{Ann.Rev.Nucl.Part.Sci.} \textbf{\bibinfo{volume}{63}},
  \bibinfo{pages}{123} (\bibinfo{year}{2013}), \eprint{1301.2826}.

\bibitem[{\citenamefont{Petersen}(2014)}]{Petersen:2014yqa}
\bibinfo{author}{\bibfnamefont{H.}~\bibnamefont{Petersen}},
  \bibinfo{journal}{J.Phys.} \textbf{\bibinfo{volume}{G41}},
  \bibinfo{pages}{124005} (\bibinfo{year}{2014}), \eprint{1404.1763}.

\bibitem[{\citenamefont{Hongo et~al.}(2013)\citenamefont{Hongo, Hirono, and
  Hirano}}]{Hongo:2013cqa}
\bibinfo{author}{\bibfnamefont{M.}~\bibnamefont{Hongo}},
  \bibinfo{author}{\bibfnamefont{Y.}~\bibnamefont{Hirono}}, \bibnamefont{and}
  \bibinfo{author}{\bibfnamefont{T.}~\bibnamefont{Hirano}}
  (\bibinfo{year}{2013}), \eprint{1309.2823}.

\bibitem[{\citenamefont{Voloshin}(2004)}]{Voloshin:2004vk}
\bibinfo{author}{\bibfnamefont{S.~A.} \bibnamefont{Voloshin}},
  \bibinfo{journal}{Phys.Rev.} \textbf{\bibinfo{volume}{C70}},
  \bibinfo{pages}{057901} (\bibinfo{year}{2004}), \eprint{hep-ph/0406311}.

\bibitem[{\citenamefont{Cooper and Frye}(1974)}]{Cooper:1974mv}
\bibinfo{author}{\bibfnamefont{F.}~\bibnamefont{Cooper}} \bibnamefont{and}
  \bibinfo{author}{\bibfnamefont{G.}~\bibnamefont{Frye}},
  \bibinfo{journal}{Phys.Rev.} \textbf{\bibinfo{volume}{D10}},
  \bibinfo{pages}{186} (\bibinfo{year}{1974}).

\bibitem[{\citenamefont{Alver et~al.}(2008)\citenamefont{Alver, Baker,
  Loizides, and Steinberg}}]{Alver:2008aq}
\bibinfo{author}{\bibfnamefont{B.}~\bibnamefont{Alver}},
  \bibinfo{author}{\bibfnamefont{M.}~\bibnamefont{Baker}},
  \bibinfo{author}{\bibfnamefont{C.}~\bibnamefont{Loizides}}, \bibnamefont{and}
  \bibinfo{author}{\bibfnamefont{P.}~\bibnamefont{Steinberg}}
  (\bibinfo{year}{2008}), \eprint{0805.4411}.

\bibitem[{\citenamefont{Kharzeev et~al.}(2002)\citenamefont{Kharzeev, Krasnitz,
  and Venugopalan}}]{Kharzeev:2001ev}
\bibinfo{author}{\bibfnamefont{D.}~\bibnamefont{Kharzeev}},
  \bibinfo{author}{\bibfnamefont{A.}~\bibnamefont{Krasnitz}}, \bibnamefont{and}
  \bibinfo{author}{\bibfnamefont{R.}~\bibnamefont{Venugopalan}},
  \bibinfo{journal}{Phys.Lett.} \textbf{\bibinfo{volume}{B545}},
  \bibinfo{pages}{298} (\bibinfo{year}{2002}), \eprint{hep-ph/0109253}.

\bibitem[{\citenamefont{Lappi and McLerran}(2006)}]{Lappi:2006fp}
\bibinfo{author}{\bibfnamefont{T.}~\bibnamefont{Lappi}} \bibnamefont{and}
  \bibinfo{author}{\bibfnamefont{L.}~\bibnamefont{McLerran}},
  \bibinfo{journal}{Nucl.Phys.} \textbf{\bibinfo{volume}{A772}},
  \bibinfo{pages}{200} (\bibinfo{year}{2006}), \eprint{hep-ph/0602189}.

\bibitem[{\citenamefont{Kharzeev and Nardi}(2001)}]{Kharzeev:2000ph}
\bibinfo{author}{\bibfnamefont{D.}~\bibnamefont{Kharzeev}} \bibnamefont{and}
  \bibinfo{author}{\bibfnamefont{M.}~\bibnamefont{Nardi}},
  \bibinfo{journal}{Phys.Lett.} \textbf{\bibinfo{volume}{B507}},
  \bibinfo{pages}{121} (\bibinfo{year}{2001}), \eprint{nucl-th/0012025}.

\bibitem[{\citenamefont{Kharzeev and Levin}(2001)}]{Kharzeev:2001gp}
\bibinfo{author}{\bibfnamefont{D.}~\bibnamefont{Kharzeev}} \bibnamefont{and}
  \bibinfo{author}{\bibfnamefont{E.}~\bibnamefont{Levin}},
  \bibinfo{journal}{Phys.Lett.} \textbf{\bibinfo{volume}{B523}},
  \bibinfo{pages}{79} (\bibinfo{year}{2001}), \eprint{nucl-th/0108006}.

\bibitem[{\citenamefont{Adamczyk et~al.}(2013)}]{Adamczyk:2013hsi}
\bibinfo{author}{\bibfnamefont{L.}~\bibnamefont{Adamczyk}} \bibnamefont{et~al.}
  (\bibinfo{collaboration}{STAR Collaboration}) (\bibinfo{year}{2013}),
  \eprint{1302.3802}.

\bibitem[{\citenamefont{Bzdak et~al.}(2011)\citenamefont{Bzdak, Koch, and
  Liao}}]{Bzdak:2010fd}
\bibinfo{author}{\bibfnamefont{A.}~\bibnamefont{Bzdak}},
  \bibinfo{author}{\bibfnamefont{V.}~\bibnamefont{Koch}}, \bibnamefont{and}
  \bibinfo{author}{\bibfnamefont{J.}~\bibnamefont{Liao}},
  \bibinfo{journal}{Phys.Rev.} \textbf{\bibinfo{volume}{C83}},
  \bibinfo{pages}{014905} (\bibinfo{year}{2011}), \eprint{1008.4919}.

\bibitem[{\citenamefont{Pratt}(2010)}]{Pratt:2010gy}
\bibinfo{author}{\bibfnamefont{S.}~\bibnamefont{Pratt}} (\bibinfo{year}{2010}),
  \eprint{1002.1758}.

\bibitem[{\citenamefont{Schlichting and Pratt}(2010)}]{Schlichting:2010na}
\bibinfo{author}{\bibfnamefont{S.}~\bibnamefont{Schlichting}} \bibnamefont{and}
  \bibinfo{author}{\bibfnamefont{S.}~\bibnamefont{Pratt}}
  (\bibinfo{year}{2010}), \eprint{1005.5341}.

\bibitem[{\citenamefont{Wang}(2010)}]{Wang:2009kd}
\bibinfo{author}{\bibfnamefont{F.}~\bibnamefont{Wang}},
  \bibinfo{journal}{Phys.Rev.} \textbf{\bibinfo{volume}{C81}},
  \bibinfo{pages}{064902} (\bibinfo{year}{2010}), \eprint{0911.1482}.

\bibitem[{\citenamefont{Bzdak and Skokov}(2012)}]{Bzdak:2011yy}
\bibinfo{author}{\bibfnamefont{A.}~\bibnamefont{Bzdak}} \bibnamefont{and}
  \bibinfo{author}{\bibfnamefont{V.}~\bibnamefont{Skokov}},
  \bibinfo{journal}{Phys.Lett.} \textbf{\bibinfo{volume}{B710}},
  \bibinfo{pages}{171} (\bibinfo{year}{2012}), \eprint{1111.1949}.

\bibitem[{\citenamefont{Bloczynski et~al.}(2013)\citenamefont{Bloczynski,
  Huang, Zhang, and Liao}}]{Bloczynski:2012en}
\bibinfo{author}{\bibfnamefont{J.}~\bibnamefont{Bloczynski}},
  \bibinfo{author}{\bibfnamefont{X.-G.} \bibnamefont{Huang}},
  \bibinfo{author}{\bibfnamefont{X.}~\bibnamefont{Zhang}}, \bibnamefont{and}
  \bibinfo{author}{\bibfnamefont{J.}~\bibnamefont{Liao}},
  \bibinfo{journal}{Phys.Lett.} \textbf{\bibinfo{volume}{B718}},
  \bibinfo{pages}{1529} (\bibinfo{year}{2013}), \eprint{1209.6594}.

\bibitem[{\citenamefont{Fukushima et~al.}(2010)\citenamefont{Fukushima,
  Kharzeev, and Warringa}}]{Fukushima:2010vw}
\bibinfo{author}{\bibfnamefont{K.}~\bibnamefont{Fukushima}},
  \bibinfo{author}{\bibfnamefont{D.~E.} \bibnamefont{Kharzeev}},
  \bibnamefont{and} \bibinfo{author}{\bibfnamefont{H.~J.}
  \bibnamefont{Warringa}}, \bibinfo{journal}{Phys.Rev.Lett.}
  \textbf{\bibinfo{volume}{104}}, \bibinfo{pages}{212001}
  (\bibinfo{year}{2010}), \eprint{1002.2495}.

\end{thebibliography}

\appendix
\end{document}